\documentclass{book}
\usepackage{vchstyle}
\usepackage{amssymb}
\begin{document}
\entry[Solar Neutrinos]{Solar
Neutrinos\protect}{John N. Bahcall}\index{Astrophysics!Solar Neutrinos|textbf}%
How does the Sun shine? How well do we understand the evolution
and ages of stars? Does the neutrino have a mass? Can a neutrino
change its lepton number in flight? Are there weak interactions
beyond those described by the standard model of particle physics?
These are some of the questions that motivate the study of solar
neutrinos.

A neutrino is a weakly interacting particle that travels at
essentially the speed of light and has an intrinsic angular
momentum of $\frac{1}{2}$ unit ($\hbar/2$). Neutrinos are produced on Earth
by natural radioactivity, by nuclear reactors, and by high-energy
accelerators. In the Sun, neutrinos are produced by
weak interactions that occur during nuclear fusion. There
are three known types of neutrinos, each associated with a
massive lepton that experiences weak, electromagnetic, and
gravitational forces, but not strong interactions. The known
leptons are electrons, muons, and taus (in increasing order
of their rest masses).

Neutrino astronomy is difficult for the same reason it is
interesting. Because neutrinos only interact weakly with matter,
they can reach us from otherwise inaccessible regions where
photons, the traditional messengers of astronomy, are trapped.
Hence, with neutrinos we can look inside stars and examine
directly energetic physical processes that occur only in stellar
interiors. We can study the interior of the Sun or the core of a
collapsing star as it produces a supernova.

Large detectors, typically hundreds or thousands of tons
of material, are required to observe astronomical neutrinos.
These detectors must be placed deep underground to avoid
confusing the rare astronomical neutrino events with the
background interactions caused by cosmic rays and their
secondary particles, which are relatively common near the
surface of the Earth.

The nearest star, our Sun, supplies the largest known flux of
neutrinos at the Earth's surface. Every second approximately a
hundred billion solar neutrinos cross every square centimeter on
Earth. Quite naturally, the first attempt to detect astronomical
neutrinos began with an experiment to observe neutrinos produced
in the deep interior of the Sun.

For two decades, from 1968 to 1988, the only operating solar
neutrino experiment (carried out by Raymond Davis Jr. and his
colleagues and using $^{37}$Cl as a detector) yielded results in
conflict with the most accurate theoretical calculations of how
many neutrinos are produced in the Sun. This conflict between
theory and observation became known as the 'solar neutrino
problem.'

Both the theoretical and the observational results for the
chlorine experiment are expressed in terms of the solar neutrino
unit, SNU, which is the product of a characteristic calculated
solar neutrino flux (units: cm$^{-2}$\,s$^{-1}$) times a
theoretical cross section for neutrino absorption (unit: cm$^2$).
A SNU has, therefore, the units of events per target atom per
second and is chosen for convenience equal to
\un{10^{-36}}{s^{-1}}.

After two decades of critical examination of both the theory and
the experiment, both results were determined robustly. The
predicted rate for capturing solar neutrinos in a $^{37}$Cl target
is (Bahcall and Ulrich, 1988; Bahcall, 1989)
\begin{equation}\label{eq:predictedchlorine}
\mbox{Predicted rate} = (7.9 \pm 0.9)~\mbox{SNU} \, .
\end{equation}
The rate observed by R. Davis, Jr.\ (1986) and his associates in
their chlorine radiochemical detector is
\begin{equation}\label{eq:observedchlorine}
\mbox{Observed rate} = (2.1 \pm 0.3)~\mbox{SNU} \, .
\end{equation}
Both the theoretical and the experimental uncertainties are quoted
as $1\sigma$ errors.

The predictions used in Eqs.~(1) and (2) are valid for the
combined standard model, that is, the standard model of
electroweak theory (of Glashow, Weinberg, and Salam) and the
standard solar model.

Similar results to those shown in
Eq.~(\ref{eq:predictedchlorine}) and
Eq.~(\ref{eq:observedchlorine}) were obtained in 1968. The
most recent theoretical result is $8.5 \pm 1.8$ SNU (Bahcall and
Pinsonneault 2004) and the final experiment value is $2.6 \pm 0.2$
SNU (Cleveland, Daily, Davis, et al. 1998). The robustness of the
discrepancy between theory and observation stimulated the
development two generations of increasingly more powerful and
sophisticated detectors designed to find the reason why theory and
observation differ.

More is known about the Sun than about any other star and the
calculations of neutrino emission from the solar interior can be
done with relatively high precision. Solar neutrino experiments
test in a direct and rigorous way the theories of nuclear energy
generation in stellar interiors and of stellar evolution. These
tests are independent of many of the uncertainties that complicate
the comparison of the theory with observations of stellar
surfaces. For example, convection and turbulence are  important
near stellar surfaces but unimportant in the solar interior.
Hence, the solar neutrino discrepancy puzzled (and worried)
astronomers who want to use neutrino observations to understand
better how the Sun and other stars shine. Prior to June 2001 (see
discussion of SNO experiment below), the solar neutrino problem
seemed to most (but not all) physicists to indicate that
astronomers did not understand the details of the solar nuclear
fusion reactions that produce neutrinos.

Neutrinos from the Sun provide particle beams for probing the weak
interactions on distance scales that cannot be achieved with
traditional laboratory experiments. Since neutrinos from the Sun
travel astronomical distances before they reach the Earth,
experiments performed with these particle beams are sensitive to
weak-interaction phenomena that require long path lengths in order
for slow weak-interaction effects to have time to occur. The
effects of tiny neutrino masses ($\geq \un{10^{-6}$}{eV}),
unmeasurable in the laboratory, can be studied with solar
neutrinos. Moreover, neutrinos traverse an enormous amount of
matter, $10^{11} {\, \rm gm \,cm^{-2}}$, as they travel from the
center of the Sun to detectors on Earth. The huge column density
of matter that solar neutrinos traverse can give rise to 'matter
effects' on neutrino propagation that have not yet been observed
with terrestrial neutrinos.

The Sun shines by converting protons into $\alpha$ particles. The
overall reaction can be represented symbolically by the
relation
\begin{equation}\label{eq:hydrogenburning}
4p \to \alpha + 2e^+ + 2\nu_{\e} + \un{25}{MeV}\ .
\end{equation}
Protons are converted to $\alpha$ particles, positrons, and
neutrinos, with a release of about \un{25}{MeV} of thermal
energy for every four protons burned. Each conversion of
four protons to an $\alpha$ particle is known as a termination
of the chain of energy-generating reactions that
accomplishes the nuclear fusion. The thermal energy that
is supplied by nuclear fusion ultimately emerges from the
surface of the Sun as sunlight. About 600 million tons
of hydrogen are burned every second to supply the
solar luminosity. Nuclear physicists have worked for
half a century to determine the details of this
transformation.

\begin{figure}[!t]
\includegraphics[angle=-90,scale=.4]{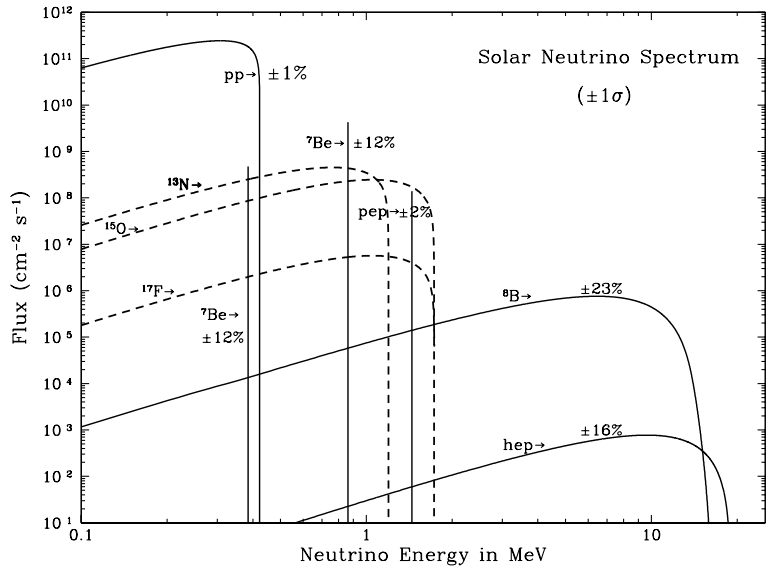}
\caption[]{Solar neutrino spectrum. This figure shows the energy
spectrum of neutrinos predicted by the standard solar
model(Bahcall and Pinsonneault 2004). The neutrino fluxes from
continuum sources (like $pp$ and $^8$B) are given in the units of
number per cm$^2$ per second per MeV at one astronomical unit. The
line fluxes ($pep$ and $^7$Be) are given in number per cm$^2$ per
sec. The spectra from the $pp$ chain (Table~\ref{tab:a20-1}) are
drawn with solid lines; the neutrino energy spectra from reactions
with carbon, nitrogen, and oxygen (CNO) isotopes are drawn with
dotted lines.} \label{fig:A20-1}
\end{figure}

The main nuclear burning reactions in the Sun are shown
in Table~1, which represents the energy-generating pp
\begin{table}
\begin{center}
\caption{The $pp$ chain in the Sun. The average number of pp
neutrinos produced per termination in the Sun is 1.85. For all
other neutrino sources, the average number of neutrinos produced
per termination is equal to the termination percentage/100.}\label{tab:a20-1}
\begin{tabular}{@{}lr@{}lr@{.}lr@{.}l@{}}
\noalign{\smallskip}
\hline
\noalign{\smallskip}
 &                            &
                                                \multicolumn{2}{c}{Termination$^a$} &
                                                          & \multicolumn{2}{c}{$\nu$ energy} \\
Reaction                             & \multicolumn{2}{l}{Number} &
                                                \multicolumn{2}{c}{(\%)} &
                                                           \multicolumn{2}{c}{(MeV)} \\
\noalign{\smallskip}
\hline
$p+p \to ^2\mathrm{H} + e^+ + \nu_e$           &  1 & \textit{a}       & \multicolumn{1}{r}{100} &       &  $\le$0 & 42 \\
or \\
$p + e^- + p \to ^2\mathrm{H}+\nu_e$           &  1 & \textit{b (pep)} &   0 & 4     &       1 & 44 \\
$^2\mathrm{H} + p \to  ^3\mathrm{He} + \gamma$ &  2 &                  & \multicolumn{1}{r}{100} &   \\
$^3\mathrm{He} + ^4\mathrm{He} \to \alpha+2p$  &  3 &                  &  \multicolumn{1}{r}{85} &   \\
or \\
$^3\mathrm{He}+^4\mathrm{He}\to^7\mathrm{Be}+\gamma$ & 4 &             &  \multicolumn{1}{r}{15} \\
~~~$^7\mathrm{Be} + e^- \to ^7Li + \nu_e$         &  5 &                &  \multicolumn{1}{r}{15} &       & (90\%) 0 & 86 \\
\multicolumn{5}{l}{\space}                                                           & (10\%) 0 & 38 \\
~~~$^7\mathrm{Li} + p \to 2\alpha$                &  6 &                &  \multicolumn{1}{r}{15} & \\
or \\
~~~$^7\mathrm{Be} + p \to ^8\mathrm{B} + \gamma$    &  7 &                &   0 & 02 \\
~~~$^8\mathrm{B}\to^8\mathrm{Be}^\ast+e^+ + \nu_e$  &  8 &                &   0 & 02    & \multicolumn{1}{r}{$<15$}    & \\
~~~$^8\mathrm{Be}^\ast \to 2\alpha$                 &  9 &                &   0 & 02 \\
or \\
$^3\mathrm{He}+p\to^4\mathrm{He}+e^+ + \nu_e$    & 10 &                &   0 & 00002 & $\le18$  & 77 \\
\noalign{\smallskip}
\hline
\end{tabular}\\
\end{center}
$^a$ The termination percentage is the fraction of terminations of
the $pp$ chain, $4p\to\alpha+2e^+ + 2\nu_e$, in which each
reaction occurs. The results are averaged over the model of the
current Sun. Since in essentially all terminations at least one
$pp$ neutrino is produced and in a few terminations one $pp$ and
one $pep$ neutrino are created, the total of $pp$ and $pep$
terminations exceeds 100\%.
\end{table}
chain. This table also indicates the relative frequency with which
each reaction occurs in the standard solar model. For simplicity,
we do not include in Table~\ref{tab:a20-1} nuclear reactions that
involve isotopes of carbon, nitrogen, and oxygen (CNO reactions).
The CNO reactions contribute only about 1\% of the solar
luminosity and relatively small neutrino fluxes (see
Figure~\ref{fig:A20-1}).

The fundamental reaction in the solar energy-generating process is
the proton-proton ($pp$) reaction, reaction 1a of
Table~\ref{tab:a20-1}, which produces the great majority of solar
neutrinos. However, these $p-p$ neutrinos have energies below the
detection thresholds for the $^{37}$Cl detector and all other
solar neutrino experiments carried out so far except radiochemical
experiments that use $^{71}$Ga as a detector (see below).

Most of the predicted capture rate in the $^{37}$Cl experiment
comes from the rare termination in which $^7$Be captures a proton
to form radioactive $^8$B (Bahcall 1964, see reaction~7 of
Table~\ref{tab:a20-1}). The $^8$B decays to unstable $^8$Be,
ultimately producing two $\alpha$ particles, a positron, and a
neutrino. The neutrinos from $^8$B decay have a maximum energy of
less than \un{15}{MeV}. Although the reactions involving $^8$B
occur only once in every 5000 terminations of the $pp$ chain, the
predicted event rates for the $^{37}$Cl, Kamiokande,
Super-Kamiokande, and SNO experiments are dominated by this rare
mode.

The neutrino energy spectrum predicted by the standard solar model
is shown in Figure~\ref{fig:A20-1}, where contributions from both
line and
\begin{table}
\begin{center}
\caption{Calculated Solar Neutrino Fluxes and $1\sigma$
Uncertainties }\label{tab:a20-2}
\begin{tabular}{@{}r@{\hspace{2em}}l@{}}
\noalign{\smallskip}
\hline
\noalign{\smallskip}
         & \multicolumn{1}{c}{Flux} \\
Source   & \multicolumn{1}{c}{(\un{10^{10}}{cm^{-2}\,s^{-1}})} \\
\noalign{\smallskip}
\hline
\noalign{\smallskip}
$pp$     & $5.9(1 \pm 0.01)$ \\
$pep$    & $0.014 (1 \pm 0.02)$ \\
he$p$    & $8  (1 \pm 0.2)\times 10^{-7}$ \\
$^7$Be   & $0.49 (1 \pm 0.12)$ \\
$^8$B    & $5.8 \times 10^{-4} (1 \pm 0.23)$ \\
$^{13}$N & $0.06 (1 \pm 0.4)$ \\
$^{15}$O & $0.05 (1 \pm 0.4)$ \\
$^{17}$F & $6 (1 \pm 0.4) \times 10^{-4} $ \\
\noalign{\smallskip}
\hline
\end{tabular}
\end{center}
\end{table}
continuum sources are included.

The solar neutrino fluxes at the Earth's surface that are
calculated from the most recent standard solar model (Bahcall and
Pinsonneault 2004) are shown in Table~2. The $1\sigma$
uncertainties in the
calculated neutrino fluxes are also shown in Table~2.

The beautiful $^{37}$Cl experiment of Davis and his collaborators
(Davis 1978, Cleveland et al. 1998) was for two decades the only
operating solar neutrino detector. The reaction that was used for
the detection of the neutrinos is
\begin{equation}\label{eq:nuabsorptionbychlorine}
\nu_e + ^{37}\mbox{Cl} \to e^- + ^{37}\mbox{Ar}\ ,
\end{equation}
which has a threshold energy of \un{0.8}{MeV}. The target was a
tank containing $10^5$ gallons of C$_2$Cl$_4$ (perchloroethylene,
a cleaning fluid), deep in the Homestake Gold Mine in Lead, South
Dakota. The underground location was necessary in order to avoid
background events from cosmic rays. Every few months, for almost
three decades, Davis and his collaborators extracted a small
sample of $^{37}$Ar, typically of order 15~atoms, out of the total
of more than $10^{30}$ atoms in the tank. The $^{37}$Ar produced
in the tank is separated chemically from the C$_2$Cl$_4$,
purified, and counted in low-background proportional counters. The
typical background counting rate for the counters corresponds to
about one radioactive decay of an $^{37}$Ar nucleus a month!
Experiments have been performed to show that $^{37}$Ar produced in
the tank is extracted with more than 90\% efficiency.

The existence of the solar neutrino problem (see
Eq.~\ref{eq:predictedchlorine} and Eq.~\ref{eq:observedchlorine})
sparked an intense debate about the origin of the problem. More
importantly, the problem stimulated the construction and operation
of five sophisticated new solar neutrino observatories.  These
observatories are: Kamiokande (a water Cherenkov detector of
neutrino-electron scattering in Japan), SAGE and GALLEX
(radiochemical detectors, in Russia and in Italy,  that observe
neutrino absorption by $^{71}$Ga), Super-Kamiokande (a much larger
version of the original Kamiokande water Cherenkov detector), and
the SNO detector (which detects neutrinos using heavy water,
$^2$H$_2$O).

The Kamiokande experiment (Kamiokande Collaboration 1996), located
in the Japanese Alps, detected Cherenkov light emitted by
electrons that are scattered in the forward direction by solar
neutrinos. The reaction by which the neutrinos are observed is
\begin{equation}\label{eq:electronscattering}
\nu + e \to \nu' + e'\ ,
\end{equation}
where the primes on the outgoing particle symbols indicate that
the momentum and energy of each particle can be changed by the
scattering interactions.  With techniques that have been developed
so far, only  the higher-energy neutrinos (\un{>5}{MeV}, i.\,e.{},
$^8$B and hep neutrinos only)  can be observed by
neutrino-electron scattering.

A much larger and more sensitive version of the Kamiokande
experiment , known as Super-Kamiokande (Super-Kamiokande
Collaboration 1998, 2001), first published new precision data in
1998. Neutrino--electron scattering experiments furnish
information about the incident neutrino energy spectrum (from
measurements of the recoil energies of the scattered electrons),
determine the direction from which the neutrinos arrive, and
record the precise time of each event. Super-Kamiokande detected
so many neutrino events (about 15 events per day, more than 5000
in total) that it inaugurated an era of precision measurements of
multiple aspects of solar neutrino interactions.

Two radiochemical solar neutrino experiments using $^{71}$Ga were
performed, one by a primarily Western European collaboration [with
U.S. and Israeli participation (see GALLEX/GNO collaboration
1992,1999, 2000)] (GALLEX) and the second by a group working in
Russia (under conditions of hardship that sometimes required
exceptional ingenuity and even heroism) with US participation
(SAGE, see SAGE Collaboration 1994, 2002. The GALLEX collaboration
used 30 tons of gallium in an aqueous solution; the detector is
located in the Gran Sasso National Laboratory in Italy. The Soviet
experiment uses about 60~tons of gallium metal as a detector in a
solar neutrino laboratory constructed underneath a high mountain
in the Baksan Valley in the Caucasus Mountains of the Soviet
Union. The amount of detector material used in each of these
experiments is impressive considering that, at the time the
experimental techniques were developed, the total world production
of gallium was only 10~tons per year!

The gallium experiments provide unique information about the most
common nuclear reaction fusion reaction in the Sun, the $p-p$
reaction (see reaction 1a of Table~\ref{tab:a20-1}).
 The absorption reaction by which neutrinos are detected with gallium is
\begin{equation}\label{eq:gallium}
\nu_e + ^{71}\mbox{Ga} \to e^- + ^{71}\mbox{Ge}\ .
\end{equation}
The germanium atoms are removed chemically from the gallium and
the radioactive decays of $^{71}$Ge are measured in small
proportional counters. The threshold for absorption of neutrinos
by $^{71}$Ga is \un{0.23}{MeV}, which is well below the maximum
energy of the $p-p$ neutrinos. The independent GALLEX/GNO and SAGE
experiments yield results that are in good agreement with each
other. At this writing, no other solar neutrino experiment has a
demonstrated capability to detect the low-energy neutrinos from
the basic $p-p$ reaction, although several detectors are being
developed that could observe electrons produced by
neutrino-electron scattering or by absorption of $p-p$ neutrinos.

The Sudbury Solar Neutrino Observatory (see SNO Collaboration
2001,2002, 2004) is a powerful 1-kiloton heavy water (D$_2$O)
experiment that is located in an INCO nickel mine near Sudbury,
Ontario (Canada). The deuterium (denoted by $D$ or by $^2$H)
experiment is a collaboration between Canadian, American, and
British scientists. Like the Kamiokande and Super-Kamiokande
detectors, the SNO deuterium detector measures the energy and
direction of recoil electrons by observing their Cherenkov light
with photomultipliers. Thus SNO can also observe neutrino-electron
scattering, see Eq. (\ref{eq:electronscattering}).

More importantly, SNO can observe two unique reactions.  The first
reaction detects only electron type neutrinos and can be written

\begin{equation}\label{eq:deuteriumnucapture} \nu_e + D
\rightarrow e^- + {\rm p + p} \, .
\end{equation}
The second reaction is equally sensitive to neutrinos of all
types, $\nu_e$, $\nu_\mu$ and $\nu_\tau$, and can be written
\begin{equation}\label{eq:deuteriumnc} \nu + D
\rightarrow \nu' + {\rm n + p} \, .
\end{equation}
The SNO detector can measure the all-neutrino reaction,
Eq. (\ref{eq:deuteriumnc}), (also called a `neutral current'
reaction) in several different ways.

\begin{figure}[!htb]
\includegraphics[angle=-90,scale=.47]{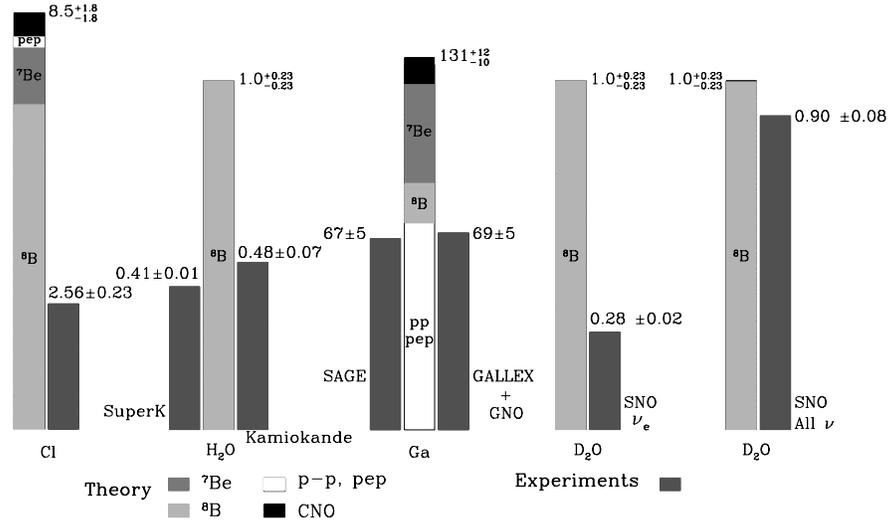}
\caption[]{Comparison of measured rates and standard-model
predictions for seven solar neutrino experiments.  The unit for
the radiochemical experiments (chlorine and gallium) is SNU
($10^{-36}$ interactions per target atom per sec); the unit for
the water-Cerenkov experiments (Kamiokande, Super-Kamiokande, and
SNO) is the rate predicted by the standard solar model plus
standard electroweak theory.  \label{fig:theoryvsexp}}
\end{figure}

Figure~\ref{fig:theoryvsexp} compares the  rates measured in all
seven of the solar neutrino experiments with the rates predicted
by the combined standard model: the standard solar model plus the
standard model of electroweak interactions. With the exception of
the neutral-current detection of SNO, all of the measurements
disagree with the predictions of the combined standard solar and
particle physics model.

The Kamiokande, Super-Kamiokande, and SNO  experiments are
sensitive to $^8$B and hep neutrinos, but the other solar
neutrinos that are shown in Figure~\ref{fig:A20-1} are below the
experimental energy thresholds.  The thresholds are set at several
MeV in order to avoid numerous lower-energy background events.
Only the gallium experiments are sensitive to the fundamental
$p-p$ neutrinos and only the gallium and chlorine experiments are
sensitive to the neutrinos from $^7$Be and from the CNO sources of
neutrinos ($^{13}$N, $^{15}$O, and $^{17}$F).


 Neutrino
absorption, exemplified by reactions
Eq. (\ref{eq:nuabsorptionbychlorine}), Eq.~(\ref{eq:gallium}), and
Eq.~(\ref{eq:deuteriumnucapture}), is sensitive only to
electron-type neutrinos, $\nu_e$, whose type (flavor) is unchanged
in transit to the Earth. For neutrino--electron scattering,
Eq.~(\ref{eq:electronscattering}), the cross section for $\nu_\mu$
or $\nu_\tau$ at the energies of interest is about one-seventh the
cross section for $\nu_e$. Neutrino--electron scattering is
primarily sensitive to $\nu_e$ but has a small sensitivity to
$\nu_\mu$ and $\nu_\tau$. The SNO experiment includes a detection
mode that is equally sensitive to all three types of neutrinos,
Eq.~(\ref{eq:deuteriumnc}). In this neutral-current mode,
deuterium nuclei are disintegrated into their constituent neutrons
and protons without changing the charge of the nucleons. The
measurement of the neutral-current disintegration of deuterium
provides a determination of the total flux of solar neutrinos
above the energy threshold, about $2.2$ MeV, for the reaction
shown in Eq.~(\ref{eq:deuteriumnc}).

On June 18th, 2001 at about 12 noon EDT the SNO collaboration
announced the first scientific results of their epochal
experiment. Combining the SNO measurements of $\nu_e$
(Eq.~\ref{eq:deuteriumnucapture}) from $^8$B neutrinos produced in
the Sun with the precise Super-Kamiokande measurement of
neutrino-electron scattering (Eq.~\ref{eq:electronscattering}),
the SNO collaboration  solved the 33 year old solar neutrino
problem. About two-thirds of the $^8$B $\nu_e$ produced in the Sun
are transformed into the more difficult to detect $\nu_\mu$ and
$\nu_\tau$ on their way from the center of the Sun to detectors on
Earth. Moreover, the total number of neutrinos of all types
($\nu_e$, $\nu_\mu$ and $\nu_\tau$) is equal, within the
uncertainties, to the value predicted by the standard solar model.

The fact that most of the neutrinos that come to us from the Sun
are transformed in flight from $\nu_e$ to $\nu_\mu$ and $\nu_\tau$
explains why the radiochemical experiments, chlorine and gallium
see less than the predicted total number of neutrinos. The
radiochemical experiments only detect $\nu_e$. The metamorphosis
from $\nu_e$ to $\nu_\mu$ and $\nu_\tau$ also explains why the
neutrino-electron scattering experiments, Kamiokande and
Super-Kamiokande, see a deficit of neutrinos. The
neutrino-electron scattering experiments are primarily, but not
entirely, sensitive to $\nu_e$.

The SNO and Super-Kamiokande measurements together established two
extraordinarily important conclusions. 1) Physics not included in
the standard model of particle physics occurs. Neutrinos change
their type. 2) The neutrino measurements confirm the theoretical
model of how the Sun shines. The measured flux of neutrinos from
$^8$B beta-decay, which depends approximately on the 25th power of
the central temperature of the Sun, is in good agreement with the
theoretical calculations.

In short, the solar neutrino experiments showed that the standard
model of particle physics is incomplete and the standard solar
model is vindicated.

Subsequent measurements by the SNO and other solar neutrino
experimental collaborations have confirmed and refined the
original inferences announced in June, 2001.

Let's step back in time for a moment to establish the theoretical
particle physics context. The physics community was electrified in
1985 when an elegant theoretical solution for the solar neutrino
problem was proposed that is consistent with expectations from
Grand Unified Theories (GUT)  of neutrino mass. According to this
solution, a $\nu_e$ created in the solar interior is almost
completely converted into $\nu_\mu$ or $\nu_\tau$ as the neutrino
passes through the Sun. This conversion reflects the enhancement
by the matter in the Sun of the probability that a neutrino of an
electron type oscillates into a neutrino of a different type; it
is universally referred to as the Mikheyev--Smirnov--Wolfenstein
(MSW) effect in honor of its discoverers.

In order for the MSW effect to occur, the flavor eigenstates
$\nu_e$, $\nu_\mu$ and $\nu_\tau$ must be different from the mass
eigenstates. The flavor eigenstates are created in weak decays and
have weak interactions with their associated charged leptons
(electron, muon, and tau) that can be written in a simple
(diagonal) form. The mass eigenstates, which have diagonal mass
matrices, are the states in which neutrinos propagate in a vacuum.
The mass eigenstates are often denoted by $\nu_1$, $\nu_2$, and
$\nu_3$. For a simplified description in terms of two eigenstates,
the relation between flavor and mass eigenstates in vacuum is
described by a single mixing angle $\theta_{12}$, where
$\tan\theta_{12}$ is the relative amplitude of $\nu_2$ and $\nu_1$
in the $\nu_e$ wave function ($\nu_e = \cos\theta_{12}\nu_1 +
\sin\theta_{12}\nu_2$). The difference in the squares of the
masses of the two neutrinos is denoted by $\Delta m^2_{21} =
m^2_{2} - m^2_{1}$.

All of the available data on solar, atmospheric, and reactor
neutrino masses are consistent with an MSW description of neutrino
propagation. A recent determination of neutrino parameters using
all the available data yields (Bahcall, Gonzalez-Garcia, and
Pe\~na-Garay 2004):

\begin{equation}
\Delta m^2_{21} ~=~ 8.2^{+0.3}_{-0.3}\times 10^{-5}\, {\rm eV^2},
\label{eq:massdifference}
\end{equation}
and
\begin{equation}
\tan^2\theta_{12} ~=~ 0.39^{+0.05}_{-0.04} \, .
\label{eq:theta12experimental}
\end{equation}
The same solution analyzing all of the data yields the  values
given below for the total flux of pp neutrinos, $\phi(pp)$,  and
the total flux of $^8$B neutrinos, $\phi(^8{\rm B})$,  both
expressed in terms of the values predicted by the standard solar
model.
\begin{equation}
\phi(pp)~=~1.01 \pm 0.02 ({\rm experimental}) \pm 0.01 ({\rm
theory})\, , \label{eq:ppflux}
\end{equation}

\begin{equation}
\phi(^8{\rm B}~=~ 0.87 \pm 0.04 ({\rm experimental}) \pm 0.23
({\rm theory}) \, . \label{eq:8Bflux}
\end{equation}
The uncertainties indicated in
Eq.~(\ref{eq:massdifference})-Eq.~(\ref{eq:8Bflux}) are
$\pm 1\sigma$ uncertainties.

\begin{bibliography}
\item J. N. Bahcall, \textit{Phys. \ Rev. \ Lett.}\ \textbf{12},
300 (1964). Showed that chlorine detection rate was expected to be
dominated by rare $^8$B neutrinos and provided theoretical
motivation for the chlorine experiment.
\item J. N. Bahcall and M. H. Pinsonneault, \textit {Phys. \ Rev.
\ Lett.} \textbf{92}, 121301 (2004).  Solar model used for
theoretical predictions in the current article.
 \item J. N. Bahcall and R. K. Ulrich, \textit{Rev.\ Mod.\
Phys.}\ \textbf{60}, 297 (1988).
      State-of-the-art calculations of solar models, neutrino fluxes,
      and helioseismological frequencies in 1988. Solar model used for  theoretical predictions
      in the article 'Neutrinos, Astronomy' by J. N. Bahcall in the second edition (1989)
      of the Encyclopedia of Physics. Very similar to current-day solar models. (A)
\item B. T. Cleveland, T. Daily, R. Davis, Jr. et al.
\textit{Astrophysical Journal}\ \textbf{496}, 505 (1998).  An
awesome and comprehensive account of a monumental experiment, the
chlorine solar neutrino experiment. (A)
 \item J. P. Cox and R. T. Guili, \textit{Principles of
Stellar Structure}. Extended second edition by A. Weiss, W.
Hillebrandt, H. -C. Thomas, and H. Ritter.
      Cambridge Scientific Publishers, Cambridge, UK 2004. Comprehensive summary
      of the theory, updated from the classic 1968 edition by four of the leading researchers of the subject. (A)
\item R. Davis, Jr., \textit{et al}., in \textit{Solar Neutrinos and Neutrino Astronomy},
      M. L. Cherry, W. A. Fowler, and K. Lande, eds., Vol.~1, p.~1.
      American Institute of Physics, New York, 1978. The classic description
      of the experiment by the master. (A)
 \item A. S. Eddington, \textit{The Internal Constitution of the
Stars}. Cambridge
      University Press, Cambridge, 1926. A beautifully written summary
      of the early theory of stellar evolution. Chapter~8 contains
      a fascinating account of the first gropings toward understanding
      of the source of stellar energy generation. (E)
\item Gallex/GNO Collaboration, P. Anselmann et al., \textit{Phys.
\ Lett.} \ \textbf{B285}, 376 (1992); W. Hampel et al.
\textit{Phys. \ Lett.} \ \textbf{B447}, 127 (1999); M. Altmann et
al. \textit{Phys. \ Lett.} \ \textbf{B490}, 16 (2000)Fundamental
radiochemical measurement that includes contribution of the pp
solar neutrinos.
\item Kamiokande Collaboration, Y. Fukuda et al.\textit{Phys. \
Rev. \ Lett.}  \ \textbf{77}, 1683 {1996}. The second solar
neutrino experiment, which confirmed the existence of the 'solar
neutrino problem', pioneered water Cherenkov detection of solar
neutrinos, and first measured the direction and energy of
individual solar neutrino events. (A)
\item S. P. Mikheyev and A. Yu. Smirnov, \textit{Sov.\ J.\ Nucl.\
Phys.}\ \textbf{42}, 913
      (1986). \textit{Nuovo Cimento} \textbf{9C}, 17 (1986); \textit{Sov.\ Phys.\ JETP} \textbf{64}, 4
      (1986); in \textit{Proceedings of 12$^{th}$ Intl.\ Conf.\ on Neutrino Physics
      and Astrophysics}, T. Kitagaki and H. Yuta, eds., p.~177. World
      Scientific, Singapore, 1986. Epochal papers, exciting to read.
      Mikheyev and Smirnov obtained by numerical integration the
      principal results for matter oscillations in the Sun and presented
      them succinctly, together with a clear physical explanation. (A)
\item B. Pontecorvo, \textit{Sov.\ JETP} \textbf{26}, 984 (1968);
      V. Gribov and B. Pontecorvo, \textit{Phys.\ Lett.}\ \textbf{B28}, 493 (1969).
      The original papers suggesting that neutrino flavor oscillations explain the solar neutrino
      problem. Founded a subject. Revolutionary ideas
      presented with clarity and brevity. (A)
\item SAGE Collaboration, J. N. Abdurashitov et al. \textit{Phys \
Lett \ B} \textbf{328}, 234 (1994); J. N. Abdurashitov et al.
\textit{J. \ Exp. \ Theor. \ Phys} \textbf{95}, 181 (2002).
Fundamental radiochemical measurement that includes contribution
of the pp solar neutrinos. (A)
\item M. Schwarzschild, \textit{Structure and Evolution of the
Stars}. Princeton
      University Press, Princeton, NJ, 1958. Classical description of
      the theory of stellar evolution with emphasis on physical understanding.
      The clearest book ever written on the subject. (I)
\item SNO Collaboration, Q. R. Ahmad et al. \textit{Phys. \ Rev. \
Lett.} \textbf{87}, 071301 (2001); ibid \textbf{89}, 011301
(2002); ibid \textbf {92}, 181301 (2004). The historic articles by
the SNO collaboration, which convinced skeptical physicists that
neutrinos change their flavor on the way to the Earth from the
center of the Sun.
\item Super-Kamiokande Collaboration, Y. Fukuda et al.,
\textit{Phys. \ Rev. \ Lett.} \textbf{81}, 1158 (1998);
\textit{Phys. \ Rev. \ Lett.} \textbf{86}, 5651  (2001) The
epochal experiment that inaugurated the era of precision
measurements in solar neutrino research, detecting more than 5,000
solar neutrino events per year. (A)
\item L. Wolfenstein, \textit{Phys.\ Rev.\ D}, \textbf{17}, 2369
(1978); \textit{Phys.\ Rev.\ D} \textbf{20},
      2634 (1979). Presented the fundamental equations for neutrino
      propagation in matter, the basis for the MSW effect. It took
      seven years for the physics community to recognize the significance
      of Wolfenstein's brilliant insight. (A)
\end{bibliography}
\end{document}